\documentclass[12pt,english]{paper}
\usepackage{amsthm,amsmath,amssymb}
\numberwithin{equation}{section}
\usepackage[T1]{fontenc}
\usepackage[latin1]{inputenc}
\usepackage{graphicx}
\usepackage{cite}
\usepackage{slashed}
\makeatletter

\newcommand{\bibstyle@aas}{\bibpunct{(}{)}{;}{a}{}{,}}

\makeatletter

\makeatletter

\usepackage{geometry}

\makeatletter

\usepackage{epsfig}

\makeatother

\makeatother

\makeatother

\usepackage{babel}
\makeatother
\begin{document}

\title{Measurement of Upward-going Milli-charged particles at the Pierre Auger Observatory}

\author{Ye Xu}

\maketitle

\begin{flushleft}
School of Electronic, Electrical Engineering and Physics, Fujian University of Technology, Fuzhou 350118, China
\par
e-mail address: xuy@fjut.edu.cn
\end{flushleft}

\begin{abstract}
It is assumed that superheavy dark matter particles with O(EeV) mass may decay to relativistic milli-charged particles (MCPs). The upward-going MCPs passing through the Earth could be measured by the fluorescence detectors (FD) of the Pierre Auger observatory (Auger). The massless hidden photon model was taken for MCPs to interact with nuclei, so that the numbers and fluxes of expected MCPs and neutrinos may be evaluated at the FD of Auger. Based on the assumption that no events are observed at the FD of Auger in 14 years, the corresponding upper limits on MCP fluxes were calculated at 90\% C. L.. These results indicated that MCPs could be directly detected in the secondaries' energy range O(1EeV)-O(10EeV) at the FD of Auger, when $3\times10^{-13}\lesssim\epsilon^2\lesssim 10^{-3}$. And a new region of 10 MeV < $m_{MCP}$ < 1 PeV and $10^{-7}$ $\lesssim$ $\epsilon$ $\lesssim$ $3\times10^{-2}$ is ruled out in the $m_{MCP}$-$\epsilon$ plane with 14 years of Auger data.
\end{abstract}


\section{Introduction}
Dark matter (DM) is one of the important topics in particle physics, cosmology and astrophysics. Sufficient evidences for the existence of DM are provided by cosmological and astrophysical observations\cite{bergstrom,BHS}. It is found in cosmological and astrophysical observations that most (84\%) of matter in the Universe consists of nonbaryonic DM\cite{Planck2015}. DM is distributed in a halo surrounding a galaxy. This halo with a local density of 0.3 GeV/cm$^3$ is assumed and its relative speed to the Sun is 230 km/s\cite{JP}. Unfortunately, no one has found these thermal DM particles yet\cite{XENON1T,PANDAX,fermi,antares-icecube-dm-MW,icecubedm-sun,antaresdm-sun,CAST,GlueX,NGC1275,Chooz,dayabayMINOS}. So far DM has been observed only through its gravitational interactions.
\par
Milli-charged particles (MCPs), which are fermions, with a small electric charge $\epsilon e$ ($e$ is the electric charge for an electron  and $\epsilon\ll1$), are an alternative DM scenario\cite{GH,CY,FLN}. A model with a hidden gauge group U(1) is taken for MCPs to interact with nuclei. The corresponding massless hidden photon field may have a kinetic mixing to the SM photon, so that a MCP under a second unbroken "mirror" U(1)$^{\prime}$ appears to have a small coupling to the SM photon\cite{Holdom}. $\epsilon$ is also the kinetic mixing parameter between those two kinds of photons. The measurements of MCPs have been performed in cosmological and astrophysical observations, accelerator experiments, experiments for decay of ortho-positronium and Lamb shift, DM searches and so on, so that constraints on $\epsilon$ were determined by those observations. \cite{CM,DHR,DGR,JR,DCB,SLAC,Xenon,LS,OP,Lamb,SUN}.
\par
Then it is a reasonable assumption that there exist at least two DM species in the Universe. In this DM scenario, one is non-thermal and non-relativistic superheavy dark matter (SHDM) $\phi$ ($m_{\phi}\gtrsim$ EeV) generated by the early universe, based on the possibility of particle production due to time varying gravitational fields\cite{KC87,CKR98,CKR99,KT,KCR,CKRT,CCKR,KST,CGIT,FKMY,FKM}. The other is the stable light fermion, MCP $\chi$, which is ultrahigh energy (UHE) products of the decay of SHDM ($\phi\to\chi\bar{\chi}$), like the DM decay channel mentioned in ref.\cite{FR}. It is assumed that their mass is much less than that of SHDM (Here $m_{\chi}\le \displaystyle\frac{m_{\phi}}{1000}$). The bulk of present-day DM consists of SHDM. Due to the decay of long-living $\phi$ ($\tau_{\phi} \gg t_0$\cite{AMO,EIP}, $t_0\sim10^{17}$ s is the age of the Universe. Here $\tau_{\phi} \geq 10^{19}$ s), the present-day DM may also contain a very small component which is MCPs with the energy of about $\displaystyle\frac{m_\phi}{2}$. Here it is assumed that $\phi$ decays only through $\phi\to\chi\bar{\chi}$.
\par
The Pierre Auger Observatory (Auger) is a detector array for the measurement of extreme energy cosmic rays (EECRs), covering an area of 3000 km$^2$, and located outside the town of Malargue, in the Province of Mendoza, Argentina. EECRs are detected in a hybrid mode at Auger, that is it consists of about 1600 surface detectors (SD) to measure secondary particles at ground level and four fluorescence detectors (FD), each consisting of 6 optical telescopes, to measure the development of extensive air showers (EAS) in the atmosphere\cite{Auger2010}. In the present work, the Earth will be taken as a part of the detector to directly measure the UHE MCPs $\chi$ induced by the decay of SHDM $\phi$ ($\phi\to\chi\bar{\chi}$). Those upward-going MCPs can be directly measured with the FD of Auger via fluorescent photons due to the deep inelastic scattering (DIS) with nuclei in the air. This detection capability will also be discussed here. The contamination consists of the UHE astrophysical neutrinos in this detection.
\section{Flux of MCPs which reach the Earth}
Here it is considered an assumption that SHDM could only decay to UHE MCPs, not SM particles. Since the lifetime for the decay of SHDM to SM particles is strongly constrained ($\tau \geq$ O($10^{26}-10^{29}$)s) by diffuse gamma and neutrino observations\cite{EIP,MB,RKP,KKK}, $\tau_{\phi}$ has to vary from $10^{19}$ s to $10^{26}$ s in the present DM scenario.
\par
The MCP flux is composed of galactic and extragalactic components. The total flux $\psi_{\chi}=\psi_{\chi}^G+\psi_{\chi}^{EG}$, where $\psi_{\chi}^G$ and $\psi_{\chi}^{EG}$ are the galactic and extragalactic MCP fluxes, respectively. The galactic MCP flux can be expressed as\cite{BLS}:
\begin{center}
\begin{equation}
\psi_{\chi}=\int_{E_{min}}^{E_{max}}C_G\frac{dN_\chi}{dE_\chi}dE
\end{equation}
\end{center}
with
\par
\begin{center}
\begin{equation}
C_G=1.7\times10^{-8}\times\frac{10^{26}s}{\tau_{\phi}}\times\frac{1TeV}{m_{\phi}} cm^{-2}s^{-1}sr^{-1}.
\end{equation}
\end{center}
where $\displaystyle\frac{dN_{\chi}}{dE_{\chi}}=\delta(E_{\chi}-\displaystyle\frac{m_{\phi}}{2})$, and E$_{\chi}$ and N$_{\chi}$ are the energy and number of MCPs, respectively.
\par
The extragalactic MCP flux can be expressed as\cite{BGG,EIP}:
\begin{center}
\begin{equation}
\psi_{\chi}^{EG}=C^{EG}\int_{E_{min}}^{E_{max}}dE \int_0^{\infty}dz\frac{1}{\sqrt{\Omega_{\Lambda}+\Omega_m(1+z)^3}}\frac{dN_\chi}{dE_\chi}[(1+z)E_\chi]
\end{equation}
\end{center}
with
\par
\begin{center}
\begin{equation}
C^{EG}=1.4\times10^{-8}\times\frac{10^{26}s}{\tau_{\phi}}\times\frac{1TeV}{m_{\phi}} cm^{-2}s^{-1}sr^{-1}.
\end{equation}
\end{center}
where z represents the red-shift of the source, $\Omega_{\Lambda}=0.685$ and $\Omega_m=0.315$ from the PLANCK experiment\cite{Planck2015}.
\section{UHE MCP and neutrino interactions with nuclei}
In this work, the hidden photon model\cite{Holdom} is taken for MCPs to interact with nuclei via a neural current (NC) interaction mediated by the mediator generated by the kinetic mixing between the SM and massless hidden photons. There is only a well-motivated interaction allowed by SM symmetries that provide a "portal", $\displaystyle\frac{\epsilon}{2}F_{\mu\nu}F^{\prime\mu\nu}$, from the SM particles into the MCPs ($F^{\prime\mu\nu}$, $F_{\mu\nu}$ are the field strength tensor of the hidden and SM photons, respectively.). Its interaction Lagrangian can be expressed as follows:
\begin{center}
\begin{equation}
\mathcal{L} =\sum_qe_q\bar{q}\gamma^{\mu}qA_{\mu} -\frac{1}{4}F^{\prime}_{\mu\nu}F^{\prime\mu\nu}+\bar{\chi}(i\slashed{D}-m_{\chi})\chi-\frac{\epsilon}{2}F_{\mu\nu}F^{\prime\mu\nu}
\end{equation}
\end{center}
where the sum runs over quark flavors in the nucleon and $e_q$ is the electric charge of the quark. $A_{\mu}$ is the vector potential of the SM photon.  $m_{\chi}$ is the MCP's mass. $\epsilon$ is the kinetic mixing parameter between the SM and hidden photons. The covariant derivative is
\begin{center}
\begin{equation}
D_{\mu}=\partial_{\mu}-ig_{\chi}A^{\prime}_{\mu}
\end{equation}
\end{center}
where $g_{\chi}$ is the gauge coupling of the U(1)$^{\prime}$ and $A^{\prime}_{\mu}$ is the vector potential of the hidden photon.
\par
The DIS cross section of MCPs on nuclei is computed with the same model in my previous work\cite{SUN}. Since the MCP-mediator coupling is equal to $\epsilon^2\alpha$, this DIS cross section is equivalent to $\epsilon^2$ times as much as that of electrons on nuclei via a NC interaction under electromagnetism, that is
\begin{center}
\begin{equation}
\sigma_{\chi N}\approx\epsilon^2\sigma^{\gamma}_{eN}
\end{equation}
\end{center}
where $\chi$ denotes a MCP with $\epsilon e$, N is a nucleon. $\sigma^{\gamma}_{eN}$ is the cross section depending on $\gamma$ exchange between elections and nuclei. The electron-nuclei cross section under electromagnetism can be obtained by integrating over the following doubly differential cross section may be expressed in term of the structure functions as\cite{SUN}
\begin{center}
\begin{equation}
\frac{d^2\sigma^{\gamma}_{eN}}{dxdQ^2}=\frac{2\pi\alpha^2}{xQ^4}Y_+F_2^{\gamma}
\end{equation}
\end{center}
where $Q^2$ is the momentum transfer, $\alpha$ is the fine-structure constant. $Y_+=1+(1-y)^2$, the inelasticity parameter $y=\displaystyle\frac{Q^2}{2m_NE_{in}}$. $m_N$ is the nucleon mass, $E_{in}$ is the incident electron energy (also the incident MCP energy). The structure function $F_2^{\gamma}$ can be expressed in terms of the quark and anti-quark parton distribution functions.
\par
The DIS cross sections of electrons on nuclei under electromagnetism can be given in table II in ref.\cite{BDH}. Those results can be approximately expressed as a simple function in the energy range 1 EeV-1 ZeV
\begin{center}
\begin{equation}
\begin{split}
\sigma^{\gamma}_{eN}\approx&1.145\times10^{-30}+2.938\times10^{-31}\rm ln[E_{in}(GeV)]-2.082\times10^{-32}\rm ln^2[E_{in}(GeV)]\\
&+1.205\times10^{-33}\rm ln^3[E_{in}(GeV)]\\
\end{split}
\end{equation}
\end{center}
The DIS cross-section for neutrino interaction with nuclei is computed in the lab-frame and given by simple power-law forms\cite{BHM} for neutrino energies above 1 PeV:
\begin{center}
\begin{equation}
\sigma_{\nu N}(CC)=4.74\times10^{-35} cm^2 \left(\frac{E_{\nu}}{1 GeV}\right)^{0.251}
\end{equation}
\end{center}
\par
\begin{center}
\begin{equation}
\sigma_{\nu N}(NC)=1.80\times10^{-35} cm^2 \left(\frac{E_{\nu}}{1 GeV}\right)^{0.256}
\end{equation}
\end{center}
where $\sigma_{\nu N}(CC)$ and $\sigma_{\nu N}(NC)$ are the DIS cross sections for neutrino scattering on nuclei via the charge current (CC) and neutral current (NC) interactions, respectively. $E_{\nu}$ is the neutrino energy.
\par
The inelasticity parameter $y=1 - \displaystyle\frac{E_{\chi^{\prime},lepton}}{E_{in}}$ (where $E_{in}$ is the incident MCP or neutrino energy and $E_{\chi^{\prime},lepton}$ is the outgoing MCPs or lepton energy). $E_{sec}=yE_{in}$, where $E_{sec}$ is the secondaries' energy after a MCP or neutrino interaction with nuclei. The mean values of $y$ for MCPs have been computed:
\begin{center}
\begin{equation}
\left\langle y \right\rangle =\frac{1}{\sigma(E_{in})} \int^1_0 y \frac{d\sigma}{dy}(E_{in},y)dy
\end{equation}
\end{center}
The MCP and neutrino interaction lengths can be obtained by
\begin{center}
\begin{equation}
L_{\nu,\chi}=\frac{1}{N_A\rho\sigma_{\nu,\chi N}}
\end{equation}
\end{center}
where $N_A$ is the Avogadro constant, and $\rho$ is the density of matter, which MCPs and neutrinos interact with.
\section{Evaluation of the numbers of expected MCPs and neutrinos}
UHE MCPs passing through the Earth and air would interact with matter of the Earth and air. UHE hadrons would be produced by MCP interaction with atmospheric nuclei. The secondary particles originated from these hadrons would develop into an EAS. And the most dominant particles in an EAS are electrons moving through atmosphere. Ultraviolet fluorescence photons would be isotropically emitted by an electron interaction with nitrogen. Their intensity is proportional to the energy deposited in the atmosphere. A small part of them may be measured by the FD of Auger (see figure 1). Since these MCP signatures are similar to DIS of neutrinos on nuclei in the atmosphere, the FD is unable to discriminate between them. Only the geometrical analysis is used to discriminate between MCPs and neutrinos. Here there exists air under an altitude of H = 100 km.
\par
\par
The number of expected MCPs, N$_{det}$, can be expressed as the following function:
\begin{center}
\begin{equation}
\begin{split}
N_{det} = & R\times \int_T \int^{E_{max}}_{E_{min}} \int_{A_{gen}} \int^{\theta_{max}}_0 \eta^{FD}_{trg} \eta^{FD}_{sel}  \Phi_{\chi} \\
  &\times P(E,D_e,D)\frac{2\pi {R_e}^2sin(\theta)}{{D_e}^2} d\theta dS dE dt \\
 = & R\times \int_T \int^{E_{max}}_{E_{min}} \int_{A_{gen}} \eta^{FD}_{trg} \eta^{FD}_{sel} dS \int^{\theta_{max}}_0 \Phi_{\chi} \\
  &\times P(E,D_e,D)\frac{2\pi {R_e}^2sin(\theta)}{{D_e}^2} d\theta dE dt
\end{split}
\end{equation}
\end{center}
where $\theta$ is the polar angle for the Earth (see figure 1), $\theta_{max}$ is the maximum of $\theta$ and taken to be $\displaystyle\frac{2\pi}{3}$ for rejecting neutrino events from the spherical crown near Auger (the declinations of astrophysical neutrinos are almost less than 60$^{\circ}$ with the IceCube detector\cite{icecubeprl,icecube2015} and the zenith angles of neutrinos are less than 95$^{\circ}$ with the Auger detector\cite{Auger2019}). $R_e$ is the radius of the Earth and taken to be 6370 km. Compared to the distance through the Earth for MCPs and neutrinos, the size of the Auger array is so small (their ratio is at least less than 2\%) that the calculation of the solid angle is simplified by the method that the Auger region is regarded as a point. T is the lifetime of taking data for Auger and taken to be 14 years in this work. R is the duty cycle for Auger and taken to be 15\%\cite{auger_upgrade}. E is the energy of an incoming particle and varies from $E_{min}$ to $E_{max}$. dS=dx$\times$dy is the horizontal surface element. Here $\Phi_\chi=\displaystyle\frac{d\psi_\chi}{dE_{\chi}}$. $\eta^{FD}_{trg}$ and $\eta^{FD}_{sel}$ are the trigger and selection efficiencies with the FD of Auger, respectively, and assumed to be related to only the energy for an incoming particle in this evaluation\cite{ICRC2005}. $S_{eff}=\int_{A_{gen}} \eta^{FD}_{trg} \eta^{FD}_{sel} dS$ is the effective area of the FD of Auger. $A_{gen}$ are the total areas where shower events hit ground level. $S_{eff}$ can be obtained from Ref.\cite{auger2011}:
\begin{center}
\begin{equation}
\begin{aligned}
 &\mathcal{A}=\displaystyle\frac{d\mathcal{E}}{dt} =\int_{\Omega} \int_{A_{gen}} \eta^{FD}_{trg} \eta^{FD}_{sel} \eta^{SD}_{trg} \eta^{SD}_{sel} cos(\theta_z) dS d\Omega \\
 &= S_{eff}\int_{\Omega} \eta^{SD}_{trg} \eta^{SD}_{sel} cos(\theta_z) d\Omega \\
 &S_{eff}=\displaystyle\frac{\mathcal{A}}{\int_{\Omega} \eta^{SD}_{trg} \eta^{SD}_{sel} cos(\theta_z) d\Omega}
\end{aligned}
\end{equation}
\end{center}
where $\mathcal{A}$ is the aperture of the Auger detector. $\theta_z=\displaystyle\frac{\theta}{2}$ is the zenith angle at Auger, and varies from $0^{\circ}$ to $60^{\circ}$. $\Omega$ is the solid angle. $\mathcal{E}$ is the hybrid exposure of the Auger detector (see figure 11 in Ref.\cite{auger2011}). The hybrid data consists of events observed on the FD and that triggered by at least one SD detector. $\eta^{SD}_{trg}$ and $\eta^{SD}_{sel}$ are the trigger and selection efficiencies with the SD of Auger, respectively. $\eta^{SD}_{trg}$ (100\% above a few EeV) can be obtained from figure 7 in Ref.\cite{SD2010}. $\eta^{SD}_{sel}$ is almost equal to 100\%\cite{SD2010}.
\par
$P(E,D_e,D)$ is the probability that MCPs interact with air after traveling a distance between $D_e$ and $D_e+D$.
\begin{center}
\begin{equation}
\begin{split}
P(E,D_e,D)=&exp\left(-\frac{D}{L_{air}}\right)exp\left(-\frac{D_e}{L_{earth}}\right)\\
 &\times \left[1-exp\left(-\frac{D}{L_{air}}\right)\right]
\end{split}
\end{equation}
\end{center}
where $D = \displaystyle\frac{H}{cos(\displaystyle\frac{\theta}{2})}$ is the effective length in the detecting zone for the FD in the air, $D_e=\displaystyle\frac{R_e(1+cos\theta)}{cos\displaystyle\frac{\theta}{2}}$ is the distances through the Earth, and $L_{earth,air}$ are the MCP interaction lengths with the Earth and air, respectively.
\par
Assuming a differential neutrino flux of $\Phi_{\nu}=k\times E_{\nu}^{-2}$, an upper limit to the value of k at 90\% C.L. is $4.4\times 10^{-9}$ GeV $cm^{-2}s^{-1}sr^{-1}$ above 0.1 EeV\cite{Auger2019}, where $\Phi_{\nu}$ represents the per-flavor flux. After k is set to be this limit, the numbers of expected neutrinos can be roughly evaluated by the above method.

\section{Results}
The numbers of expected MCPs and neutrinos are evaluated at different incoming energies (1 EeV < E <1 ZeV) assuming 14 years of Auger data. Figure 2 shows the distributions of expected MCPs and neutrinos in the secondaries' energy range 1 EeV-100 EeV. Compared to MCPs with $\epsilon^2$=$3\times10^{-13}$, the numbers of expected neutrinos are at least smaller by 14 orders of magnitude in the energy range 1 EeV-100 EeV. So the neutrino contamination can be ignored in this measurement. The numbers of expected MCPs with $\epsilon^2$=$3\times10^{-13}$ (see the cyan dash dot dot line) and $\epsilon^2$=$10^{-3}$ (see the red solid line) can reach about 1 at energies with near 1 EeV at the FD of Auger, respectively. Figure 2 also presents MCPs with $\epsilon^2$=$10^{-12}$(see the blue dash dot line) and $10^{-11}$ (see the green dot line) could be detected below about 5 EeV and 37 EeV at the FD of Auger, respectively, when $\tau_{\phi} = 10^{19}$ s.
\par
Ref.\cite{ICRC2021} presents an analysis of upward-going showers at the FD of Auger. An observed event has been found with 14 years of Auger data, but it is likely to be a downward-going monocular event reconstructed erroneously as an upward-going shower\cite{ICRC2021}. So this analysis has not found any significant indication of upward-going MCPs and neutrinos. Thus it is a reasonable assumption that no events are observed in the measurement of upward-going MCPs and neutrinos at the FD of Auger in 14 years. The corresponding upper limit on MCP flux at 90\% C.L. was calculated with the Feldman-Cousins approach\cite{FC} (see the black solid line in figure 3). Figure 3 also presents the fluxes of expected MCPs with $\epsilon^2=3\times10^{-13}$ (cyan dash dot dot line), 10$^{-12}$ (blue dash dot line), 10$^{-11}$ (green dot line) and $1.05\times10^{-3}$ (red solid line). As shown in figure 3, however, no ranges of expected MCP flux are ruled out by that limit in the energy range 1 EeV-100 EeV with $\epsilon^2=3\times10^{-13}$, $10^{-12}$, $10^{-11}$ and $1.05\times10^{-3}$, respectively.
\section{Discussion and Conclusion}
With $\epsilon^2$ = $3\times10^{-13}$, 10$^{-12}$, 10$^{-11}$ and $1.05\times10^{-3}$, hence, the upward-going MCPs can be measured at energies below about 1.4 EeV, 5 EeV, 37 EeV and 1 EeV at the FD of Auger, respectively, when $\tau_{\phi} = 10^{19}$ s. Based on the results described above, it is a reasonable conclusion that those MCPs could be directly detected in the energy range O(1EeV)-O(10EeV) at the FD of Auger when $3\times10^{-13}<\epsilon^2<1.05\times10^{-3}$. Since these constraints are only given by the assumptions mentioned above, certainly, the experimental collaborations, like the Pierre Auger collaboration, should be encouraged to conduct an unbiased analysis with the data of Auger.
\par
Since $\Phi_{MCP}$ is proportional to $\displaystyle\frac{1}{\tau_{\phi}}$ (see equation 2.1-2.4), the above results actually depends on the lifetime of SHDM, $\tau_{\phi}$. If $\tau_{\phi}$ varies from $10^{20}$ s to $10^{21}$ s, the numbers of expected MCPs at the FD of Auger are less by from 1 to 2 orders of magnitude than that with $\tau_{\phi}=10^{19}$ s.
\par
Likewise, the upper limit for $\epsilon^2$ at 90\% C.L. can be calculated with the Feldman-Cousins approach. Figure 4 shows these limits with $\tau_{\phi}$ = $10^{19}$ s (see red solid line), $10^{20}$ s (see green solid line) and $5\times10^{20}$ s (see blue solid line), respectively. If the SHDM mass, $m_{\phi}$, is equal to 2 EeV (the corresponding MCP energy is just 1 EeV), as shown in figure 4, the region of $\epsilon^2>1.5\times10^{-14}$ (that is $\epsilon$ > $1.2\times10^{-7}$) is ruled out when $\tau_{\phi}=10^{19}$ s.
\par
The MCP mass, $m_{MCP}$, is at least taken to be less than 1 PeV, since it is assumed that the MCP mass are much less than that of SHDM, as mentioned in section 1. With $\epsilon^2$ over about $10^{-5}$, since the MCP cross-sections scattering off nuclei become so large that few MCPs can pass through the Earth and reach the detector, the detected MCPs would be significantly reduced. Considering at least a MCPs can be detected, $\epsilon^2<1.05\times10^{-3}$ (that is $\epsilon<3.2\times10^{-2}$). So the region of $1.2\times10^{-7}<\epsilon<3.2\times10^{-2}$ is ruled out in the $m_{MCP}$-$\epsilon$ plane, when $m_{MCP} < $ 1 PeV. This result is shown in figure 5. To compare to other observations on MCPs, this figure also shows the $\epsilon$ bounds from cosmological and astrophysical observations\cite{CM,DHR,DGR,JR}, accelerator and fixed-target experiments\cite{DCB,SLAC}, experiments for decay of ortho-positronium\cite{OP}, Lamb shift\cite{Lamb} and measurement of MCPs from the sun's core\cite{SUN}. A new region of 10 MeV < $m_{MCP}$ < 1 PeV and $10^{-7}\lesssim\epsilon\lesssim3\times10^{-2}$ is ruled out in the $m_{MCP}$-$\epsilon$ plane with 14 year of Auger data, as shown in figure 5.
\par
Since the decay of SHDM into MCPs would lead to extra energy injection during recombination and reionization eras in the early universe, the parameters in this DM scenario may be constrained by early universe observations. Since the lifetime of SHDM is much greater than the age of the Universe, however, $\Omega_{MCPs} h^2 \lesssim 10^{-12}\Omega_{DM}h^2$ in this scenario. Ref.\cite{DDRT} presented that the cosmological abundance of MCPs was strongly constrained by the Planck data ($\Omega_{MCPs}h^2<0.001$). Thus, the parameters in this scenario can't be constrained by the present early universe observations.
\section{Acknowledgements}
This work was supported by the National Natural Science Foundation
of China (NSFC) under the contract No. 11235006, the Science Fund of
Fujian University of Technology under the contracts No. GY-Z14061 and GY-Z13114 and the Natural Science Foundation of
Fujian Province in China under the contract No. 2015J01577.
\par

\newpage

\begin{figure}
 \centering
 \includegraphics[width=0.9\textwidth]{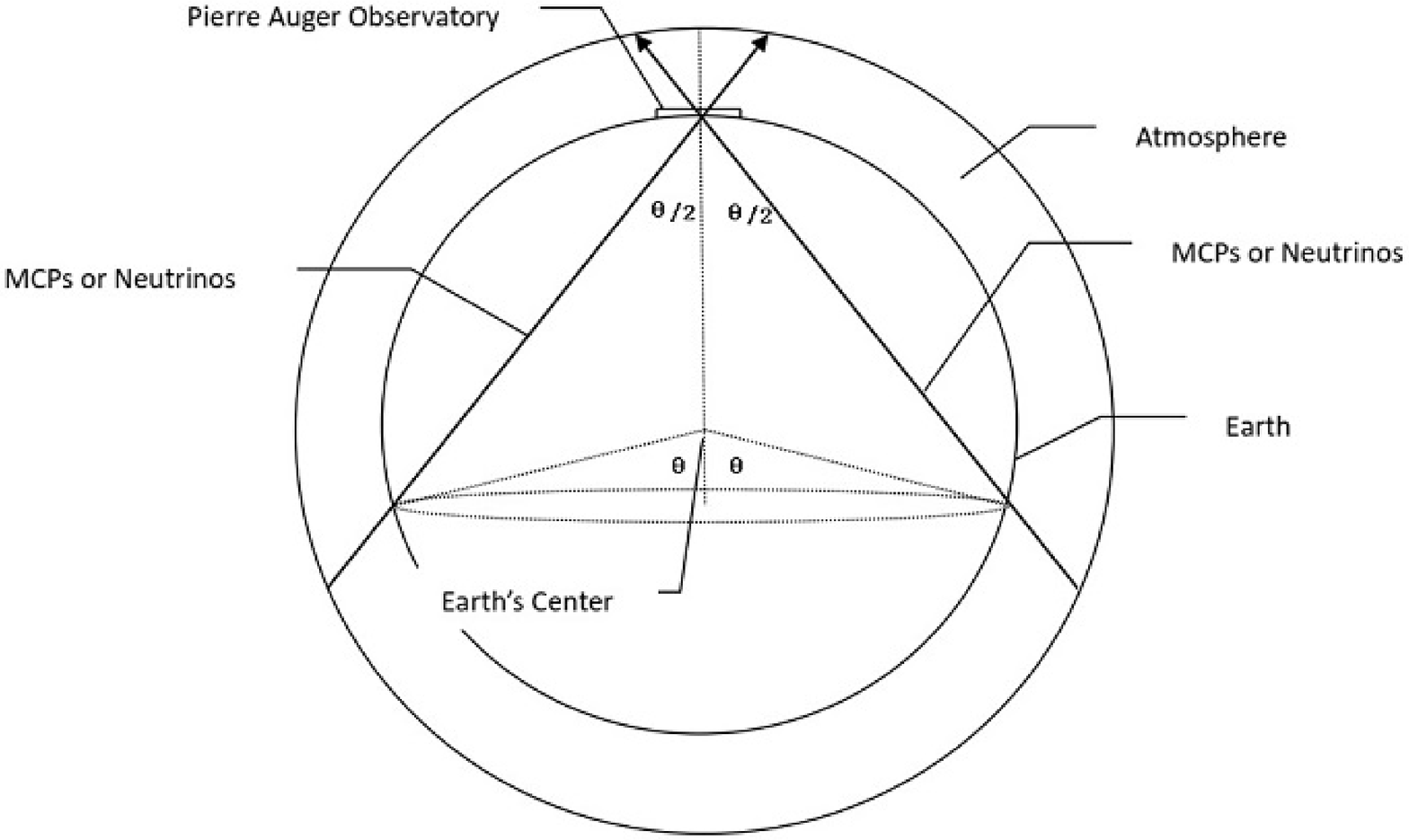}
 \caption{UHE MCPs pass through the Earth and air and can be measured by the FD of Auger via fluorescent photons due to the development of an EAS. $\theta$ is the polar angle for the Earth.}
 \label{fig:figure}
\end{figure}

\begin{figure}
 \centering
 \includegraphics[width=0.9\textwidth]{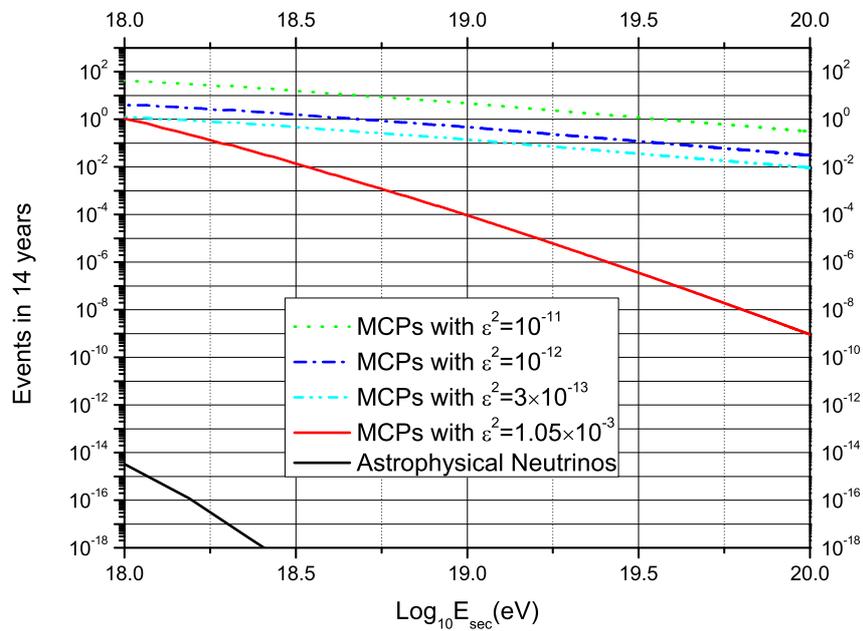}
 \caption{With the different $\epsilon^2$ (= $3\times10^{-13}$, $10^{-12}$, $10^{-11}$ and $1.05\times10^{-3}$), the numbers of expected MCPs were evaluated assuming 14 years of Auger data, respectively. The evaluation of numbers of expected neutrinos was also computed.}
 \label{fig:event}
\end{figure}

\begin{figure}
 \centering
 \includegraphics[width=0.9\textwidth]{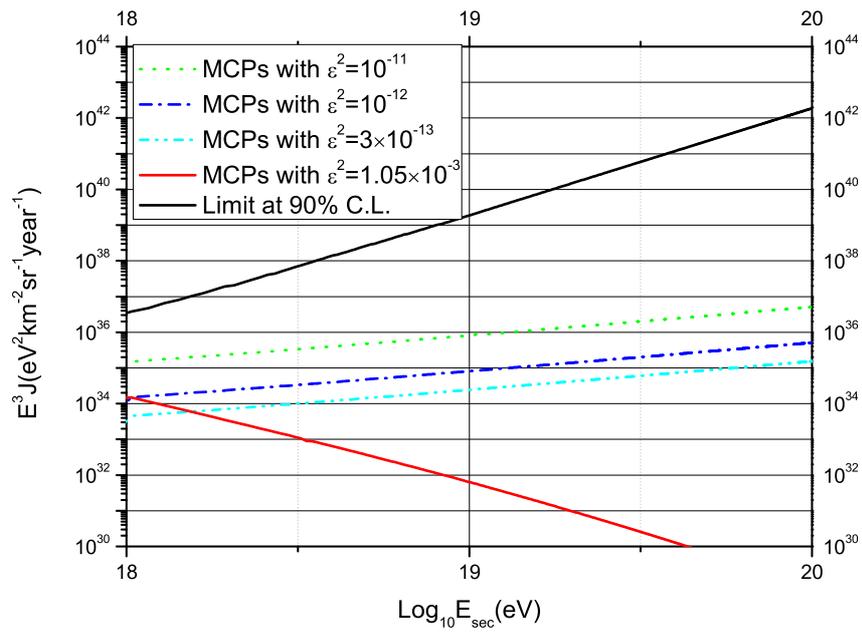}
 \caption{With the different $\epsilon^2$ (= $3\times10^{-13}$, $10^{-12}$, $10^{-11}$ and $1.05\times10^{-3}$), the fluxes of expected MCPs were estimated at the FD of Auger, respectively. Assuming no observation at Auger in 14 years, the upper limit at 90\% C.L. was also computed.}
 \label{fig:flux_1e19}
\end{figure}

\begin{figure}
 \centering
 \includegraphics[width=0.9\textwidth]{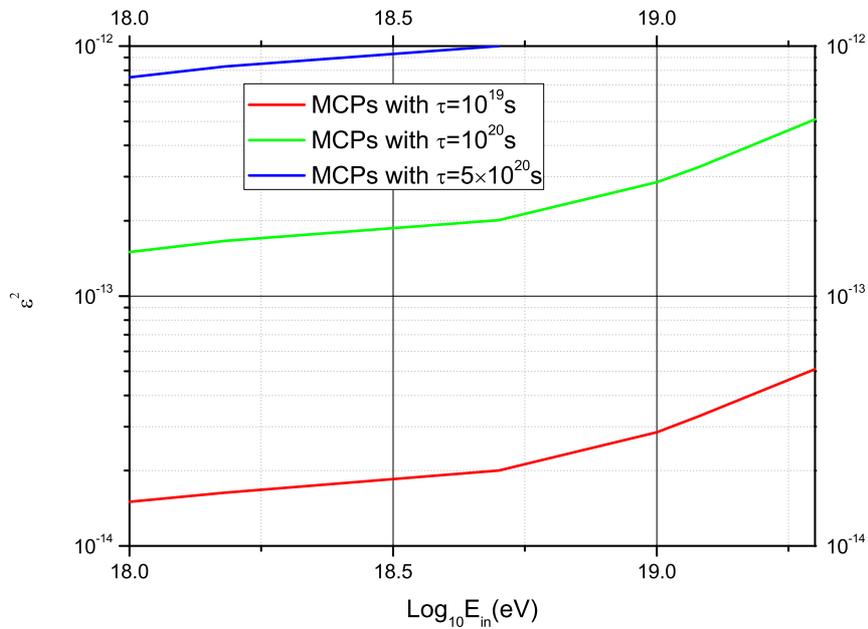}
 \caption{With the different $\tau_{\phi}$ (= $10^{19}$ s, $10^{20}$ s and $5\times10^{20}$ s), the upper limit on $\epsilon^2$ at 90\% C.L. was computed, respectively, assuming no observation at Auger in 14 years.}
 \label{fig:uplim_epsilon2}
\end{figure}

\begin{figure}
 \centering
 \includegraphics[width=0.9\textwidth]{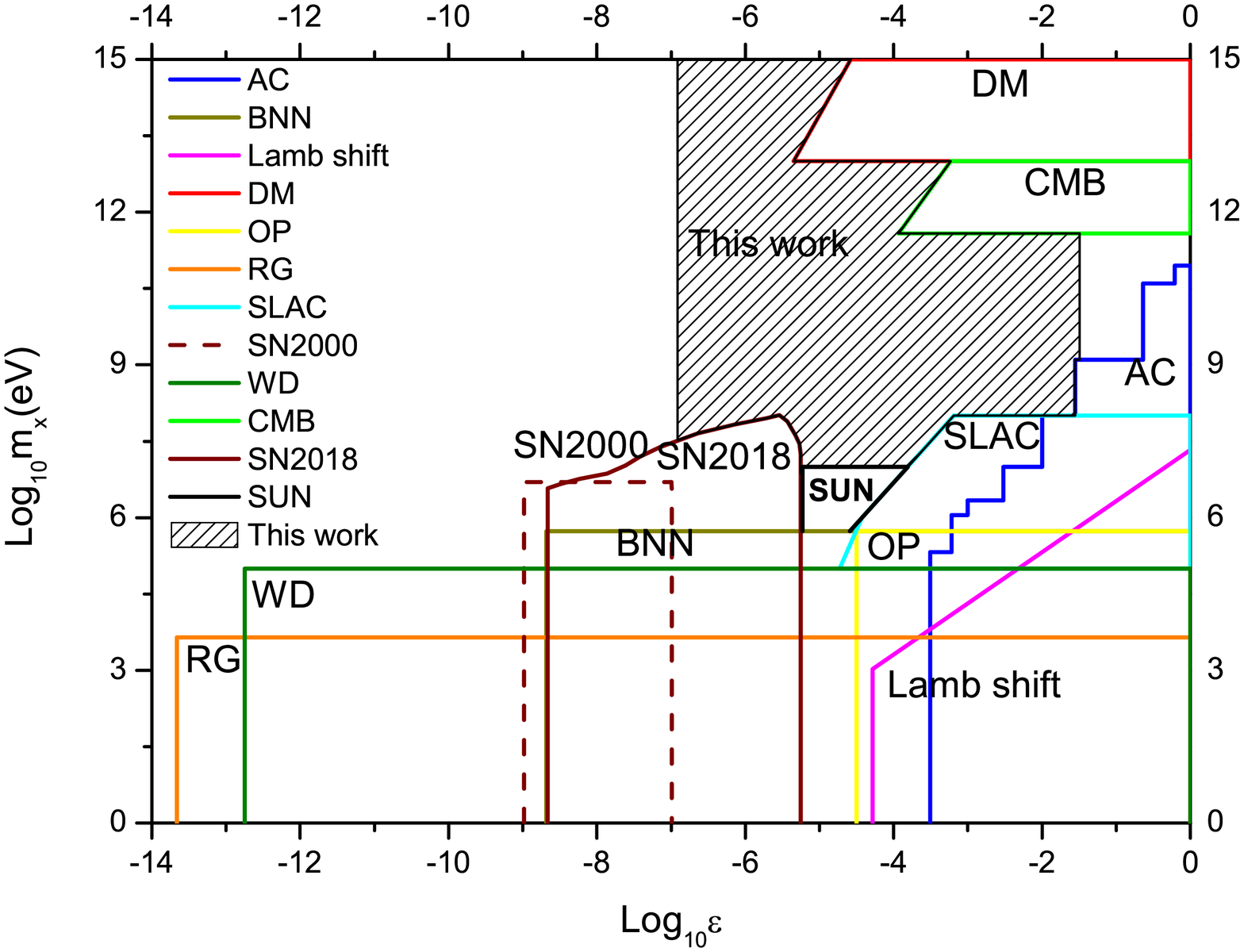}
 \caption{If $m_{\phi}$=2 EeV, with $\tau_{\phi}=10^{19}$ s, a new region (shaded region) is ruled out in the $m_{MCP}$ vs. $\epsilon$ plane, when $m_{MCP} < 1 PeV$ and $\epsilon > 1.2\times10^{-7}$ (this work). Meanwhile, the bounds from plasmon decay in red giants (RG)\cite{DHR}, plasmon decay in
white dwarfs (WD)\cite{DHR}, cooling of the Supernova 1987A (SN2000\cite{DHR}, SN2018\cite{CM}), accelerator (AC)\cite{DCB} and fixed-target experiments (SLAC)\cite{SLAC}, the Tokyo search for the invisible decay of ortho-positronium (OP)\cite{OP}, the Lamb shift\cite{Lamb}, big bang nucleosynthesis (BBN)\cite{DHR}, cosmic microwave background (CMB)\cite{DGR}, dark matter searches (DM)\cite{JR}, measurement of MCPs from the sun's core\cite{SUN} are also plotted on this figure.}
 \label{fig:epsilon_bound}
\end{figure}


\begin{thebibliography}{}
\bibitem{bergstrom}L. Bergstrom, Rept. Prog. Phys. 63,793 (2000) arXiv: hep-ph/0002126
\bibitem{BHS}G. Bertone, D. Hooper and J.Silk, Phys. Rep. 405, 279 (2005) arXiv: hep-ph/0404175
\bibitem{Planck2015}P.A.R. Abe, et al., Planck collaboration, A\&A 594, A13 (2015) arXiv:1502.01589
\bibitem{JP}J.D.Lewin, P.F.Smith, Astropart. Phys. 6, 87 (1996)
\bibitem{XENON1T} E.Aprile, et al., XENON1T Collaboration, Phys. Rev. Lett. 119, 181301 (2017)
\bibitem{PANDAX} X.Y.Cui, et al., PandaX-II Collaboration, Phys. Rev. Lett. 119, 181302 (2017)
\bibitem{fermi}M. Ackermann, et al., Fermi-LAT Collaborations, JCAP 09, 008 (2015) arXiv: 1501.05464
\bibitem{antares-icecube-dm-MW} A. Albert, M. Andre, et al., ANTARES and IceCube Collaboration, Phys. Rev. D 102, 08002 (2020) arXiv: 2003.06614
\bibitem{antaresdm-sun}S. Adrian-Matinez, et al., ANTARES Collaboration, Phys. Lett. B 759, 69-74, (2016), arXiv: 1603.02228
\bibitem{icecubedm-sun}M. G. Aartsen, et al., IceCube Collaboration, Euro. Phys. J. C 77, 146 (2017) arXiv: 1612.05949
\bibitem{CAST}V. Anastassopoulos, et al., CAST Collaboration, Nature Physics 13, 584 (2017)
\bibitem{GlueX}S. Adhikari, C.S. Akondi, GlueX Collaboration, Phys. Rev. D 105, 052007 (2022) arXiv: 2109.13439
\bibitem{NGC1275} C.S. Reynolds et al., Astrophys. J. 890, 59 (2020)
\bibitem{Chooz} T. Abrahao, et al., Double Chooz Collaboration, Eur. Phys. J. C 81, 775 (2021) arXiv: 2009.05515
\bibitem{dayabayMINOS}P. Adamson, F.P. An, et al., Daya Bay, MINOS+ Collaboration, Phys. Rev. Lett. 125, 071801 (2020) arXiv: 2002.00301
\bibitem{GH}H. Goldberg and L.J. Hall, Phys. Lett. B 174, 151 (1986)
\bibitem{CY}K.Cheung and T.C. Yuan, JHEP 0703, 120 (2007) arXiv: hep-ph/0701107
\bibitem{FLN}D. Feldman, Z. Liu and P. Nath, Phys. Rev. D 75, 115001 (2007) arXiv: hep-ph/0702123
\bibitem{Holdom}B. Holdom, Phys. Lett. B 166, 196 (1986)
\bibitem{CM}J. H. Chang, R. Essig, and S. D. McDermott, J. High Energy Phys. 09, 051 (2018)
\bibitem{DHR}S. Davidson, S. Hannestad, and G. Raffelt, J. High Energy Phys. 05, 003 (2000)
\bibitem{DGR}S. Dubovsky, D. Gorbunov and G. Rubtsov, JETP Lett. 79 (2004) arXIv: hep-ph/0311189
\bibitem{JR}J. Jaeckel and A. Ringwald, Annu. Rev. Nucl. Part. Sci. 60, 405-437 (2010)
\bibitem{DCB}S. Davidson, B. Campbell and D. Bailey, Phys. Rev. D 43, 2314 (1991)
\bibitem{SLAC}A. Prinz et al., SLAC Collaboration, Phys. Rev. Lett. 81, 1175 (1998) arXiv: hep-ex/9804008
\bibitem{Xenon}R. Essig, T. Volansky, and T.T. Yu, Phys. Rev. D 96, 043017 (2017)
\bibitem{LS}H. Liu and T. R. Slatyer, Phys. Rev. D 98, 023501 (2018)
\bibitem{OP}T. Mitsui et al., Phys. Rev. Lett. 70, 2265 (1993)
\bibitem{Lamb}S.R. Lundeen, F.M. Pipkin, Phys. Rev. Lett. 46, 232 (1981)
\bibitem{SUN} Y. Xu, JHEP 09, 055 (2022) arXiv: 2207.00178
\bibitem{KC87} M.Yu.Khlopov, V.M.Chechetkin, Sov. J. Part. Nucl 18, 267-288 (1987)
\bibitem{CKR98} D. J. H. Chung, E. W.Kolb, and A.Riotto, Phys.Rev.Lett. 81, 4048, (1998) arXiv: hep-ph/9805473
\bibitem{CKR99} D. J. H. Chung, E. W.Kolb, and A.Riotto, Phys.Rev. D59, 023501 (1998) arXiv: hep-ph/9802238
\bibitem{KT}V. Kuzmin and I. Tkachev, JETP Lett. 68, 271¨C275 (1998) arXiv: hep-ph/9802304
\bibitem{KCR} E. W. Kolb, D. J. Chung, and A. Riotto, WIMPzillas!, hep-ph/9810361
\bibitem{CKRT} D. J. H. Chung, E. W. Kolb, A. Riotto, and I. I. Tkachev, Phys. Rev. D 62, 043508 (2000) arXiv: hep-ph/9910437
\bibitem{CCKR}D. J. H. Chung, P. Crotty, E. W. Kolb, and A. Riotto,, Phys. Rev. D 64, 043503 (2001) arXiv: hep-ph/0104100
\bibitem{KST} E. W. Kolb, A. Starobinsky, and I. Tkachev, JCAP 0707, 005 (2007) arXiv: hep-th/0702143
\bibitem{CGIT} L.Covi, M.Grefe, A.Ibarra and D.Tran, JCAP 1004, 017 (2010) arXiv: 0912.3521
\bibitem{FKMY} B.Feldstein, A.Kusenko, S.Matsumoto and T. T.Yanagida Phys. Rev. D 88, 015004 (2013) arXiv: 1303.7320
\bibitem{FKM}M. A. Fedderke, E. W. Kolb, and M. Wyman, Phys. Rev. D 91, 063505 (2015) arXiv:1409.1584
\bibitem{FR}Y. Farzan and M. Rajaee, JCAP 04, 040 (2019) arXiv: 1901.11273
\bibitem{AMO}R. Aloisio, S. Matarrese and A. V. Olinto, JCAP, 1508, 024 (2015), arXiv: 1504.01319
\bibitem{EIP} A.Esmaili, A.Ibarra and O.L. Peres, JCAP, 1211, 034 (2012) arXiv:1205.5281

\bibitem{Auger2010} J. Abraham, et al., The Pierre Auger Collaboration, Phys. Lett. B 685, 239-246 (2010) arXiv: 1002.1975
\bibitem{MB} K.Murase and J.F.Beacom, JCAP 1210, 043 (2012) arXiv:1206.2595
\bibitem{RKP} C.Rott, K.Kohri and S.C.Park, Phys. Rev. D 92, 023529 (2015) arXiv:1408.4575
\bibitem{KKK}M.Kachelriess, O.E.Kalashev and M.Yu.Kuznetsov, Phys. Rev. D 98, 083016 (2018) arXiv: 1805.04500
\bibitem{BLS} Y.Bai, R.Lu and J.Salvado, JHEP, 01, 161 (2016) arXiv:1311.5864
\bibitem{BGG} A.Bhattacharya, R.Gandhi and A.Gupta, JCAP 1503, 027, (2015) arXiv:1407.3280

\bibitem{BDH} M.M. Block, L. Durand and P. Ha, Phys. Rev. D89, 094027 (2014) arXiv: 1404.4530
\bibitem{BHM} Martin M.Block, Phuoc Ha, Douglas W.McKay, Phys. Rev. D 82, 077302 (2010) arXiv: 1008.4555
\bibitem{icecubeprl}M.G. Aartsen, et al., the IceCube Collaboration, Phys. Rev. Lett., 113, 101101 (2014)
\bibitem{icecube2015}M.G. Aartsen, et al., the IceCube Collaboration, Proceedings of the 34th International Cosmic Ray Conference, Hague, Netherlands, 30 July - 6 August 2015, 1081, arXiv: 1510.05223
\bibitem{Auger2019}A.Abe, et al., the Pierre Auger Collaboration, Contributions of the Pierre Auger Collaboration to the 36th International Cosmic Ray Conference (ICRC 2019), 24 July - 1 August 2019, Madison, Wisconsin, USA, arXiv: 1909.09073
\bibitem{auger_upgrade} A. Aab, et al., the Pierre Auger Collaboration, The Pierre Auger Observatory Upgrade: Preliminary Design Report, arXiv:1604.03637
\bibitem{ICRC2005}J.A. Bellido, et al., the Pierre Auger Collaboration, Proceedings of the 29th International Cosmic Ray Conference, Pune, India, 3-10 August 2005, 101-106, arXiv: astro-ph/0507103
\bibitem{auger2011}P. Abreu, et al., the Pierre Auger Collaboration, Astroparticle Physics, 34, 368-381 (2011) arXiv: 1010.6162
\bibitem{SD2010}J. Abraham, et al., the Pierre Auger Collaboration, NIMA, 613, 29-39 (2010) arXiv: 1111.6764
\bibitem{ICRC2021}M. Mastrodicasa on behalfof the Pierre Auger Collaboration, Proceedings of the 37th International Cosmic Ray Conference, Berlin, Germany, 12-23 July 2021, 1140
\bibitem{FC}G. J. Feldman, R. D. Cousins, Phys. Rev. D 57, 3873 (1998)

\bibitem{DDRT}A.D. Dolgov, S.L. Dubovsky, G.I. Rubtsov and I.I. Tkachev, Phys. Rev. D 88, 117701 (2013) arXiv: 1310.2376


\end{thebibliography}
\end{document}